\documentclass[11pt]{article}

\input{preface/includes}
\newlength{\arrow}
\newcommand*{\goesto}[1]{\xrightarrow{\mathmakebox[\arrow]{#1}}}
\newcommand{\vin}{v_\text{in}}
\newcommand{\Vin}{V_\text{in}}
\newcommand{\vout}{v_\text{out}}
\newcommand{\Vout}{V_\text{out}}
\newcommand{\Cp}{C^+}
\newcommand{\Cm}{C^-}
\newcommand{\Sp}{S^+}
\newcommand{\Sm}{S^-}
\newcommand{\cp}{c^+}
\newcommand{\cm}{c^-}
\renewcommand{\sp}{s^+}
\newcommand{\sm}{s^-}
\newcommand{\Mp}{M^+}
\newcommand{\Mm}{M^-}
\renewcommand{\mp}{m^+}
\newcommand{\mm}{m^-}
\newcommand{\D}[1]{\frac{d#1}{dt}}

\numberwithin{equation}{section}

\newtheorem*{observation*}{Observation}

\newcommand{\savefigurename}{\figurename}

\NewEnviron{crn}{
	\begin{mdframed}[userdefinedwidth=0,
	                 innertopmargin=-18,
	                 innerrightmargin=6,
	                 innerleftmargin=6,
	                 align=center]
		\begin{align*} \BODY \end{align*}
	\end{mdframed}
}

\NewEnviron{figcrn}[2]{
	\renewcommand{\figurename}{CRN}
	\begin{figure}
		\begin{mdframed}[userdefinedwidth=0,
		                 innertopmargin=-18,
		                 innerrightmargin=6,
		                 innerleftmargin=6,
		                 align=center]
			\begin{align*} \BODY \end{align*}
		\end{mdframed}
		\caption{\label{#2}  #1}
	\end{figure}
	\renewcommand{\figurename}{\savefigurename}
}

\begin{document}

\title{
    Modulated Signals in\\Chemical Reaction Networks%
    \thanks{This research was supported in part by National Science Foundation Grants 1247051, 1545028, and 1900716.}
}

%%%%%%%%%%%%%%%%%%%%%%%%%%%%%%%%%%%%%%%%%%%%%%%%%%%%%%%%

\author[1]{Titus H. Klinge}
\affil[1]{Drake University, Des Moines, IA 50311 USA
    \texttt{titus.klinge@drake.edu}}
\author[2]{James I. Lathrop}
\affil[2]{Iowa State University, Ames, IA 50011 USA
    \texttt{jil@iastate.edu}}

\date{}

%%%%%%%%%%%%%%%%%%%%%%%%%%%%%%%%%%%%%%%%%%%%%%%%%%%%%%%%

\maketitle

    \begin{abstract}
Electrical engineering and molecular programming share many of the same mathematical foundations.
In this paper, we show how to send multiple signals through a single pair of chemical species using modulation and demodulation techniques found in electrical engineering.
Key to our construction, we provide chemical implementations of classical linear band-pass and low-pass filters with induced differential equations that are identical to their electrical engineering counterparts.
We show how to modulate \emph{arbitrary} independent input signals with different carrier frequencies for transmission through a shared medium. Specific signals in the medium can then be isolated and demodulated using band-pass and low-pass filters.
Such programmable chemical band-pass filters also offer a way to monitor chemical systems to verify that they are operating between a prescribed set of frequencies.
    % \vspace*{1em}
    % \noindent
    % \textbf{Keywords: }
    % Biomolecular automata;
    % Input/output chemical reaction networks;
    % Concentration signals; 
    % Molecular programming;
    % Robustness
\end{abstract}

    \section{Introduction}
The chemical reaction network model is commonly used to prototype nanodevices~\cite{cWinf19,jVSK20,jCHK+20}.
In particular, a chemical reaction network (CRN) models the molecular interactions of chemical species and is related to distributed models of computation such as population protocols~\cite{jAAD+06,cDE19}.
One of the most common variants of the CRN model is \emph{deterministic} and is equivalent in power to Shannon's general-purpose analog computer~\cite{jShan41,jGCB08c} and is  Turing complete~\cite{cFLB+17}.
Deterministic CRNs model the evolution of chemical species as real-valued \emph{concentrations} whose rate of change is governed by mass action kinetics~\cite{bPJ89,bFein19}.
Since distributed biochemical systems communicate via molecular concentration signals, investigation of various molecular communication techniques is ongoing~\cite{jFYE+16,jKDB+19,jBDP+20}.

In this paper we show how to multiplex multiple chemical signals on a single pair of chemical species using radio communication techniques.
We show that classical amplitude modulation and demodulation of signals on specific carrier frequencies can be accomplished with simple, small, and natural chemical reaction networks.
Our work is related to that of Cardelli, Tribastone, and Tschaikowski who recently showed that any linear electric circuit can be converted to a chemical reaction network that approximates its behavior~\cite{jCTT20}.
In contrast, our chemical low-pass and band-pass implementations yield solutions that \emph{exactly} simulate their electronic counterparts.
Moreover, their general approach requires foreknowledge of the input signal ODEs, whereas our implementations are entirely input-agnostic.

Modulation is accomplished through two reactions that multiply an input signal with its corresponding carrier frequency signal, producing a dual-rail signal using species $M^+$ and $M^-$; these two species represent the shared communication channel.
Our modulation scheme uses $2s+2$ reactions to combine the carrier signals, where $s$ is the number of signals being modulated onto the medium.
However, each modulated signal must also have a corresponding carrier frequency encoded as a chemical signal.
For each input, we use four additional reactions to create a dual-rail sinusoidal signal, tuned to its target carrier frequency using the concentration of a catalyst species.
It is important to note that our construction does not require a perfect sinusoidal wave; almost any waveform of sufficient frequency can be used.

Recovering the AM modulated signal is accomplished using a band-pass filter to isolate a particular modulated carrier frequency, followed by a rectification and low-pass filter to reconstruct the original signal.
Both the chemical low-pass and band-pass filters admit an arbitrary input, allowing the system to operate even with unknown signals.
The band-pass filter alone may be of interest for detecting if a system is oscillating in a range of frequencies.
For example, many biological and chemical systems only function with oscillations within a frequency range~\cite{bioStuff1,bioStuff2}.
Since the band-pass filter can be constructed using catalytic single rail inputs, a chemical reaction system can be devised that alarms when a target chemical species is not operating between prescribed frequencies without affecting the system.

The rest of this paper is divided as follows.
Section~\ref{sec:prelim} gives some basic information about chemical reaction networks and filters.
Section~\ref{sec:band_pass} describes the construction of a programmable band-pass filter as well as a low-pass filter.
Section~\ref{sec:modulation} shows how to modulate an arbitrary signal, transmit it via a medium (with other signals), and then demodulate the signal.
    
    %%%%%%%%%%%%%%%%%%%%%%%
\section{Preliminaries \label{sec:prelim}}
%%%%%%%%%%%%%%%%%%%%%%%

%%%%%%%%%%%%%%%%%%%%%%%%%%%%%%%%%%%%%%%
% \subsection{Chemical reaction networks}
%%%%%%%%%%%%%%%%%%%%%%%%%%%%%%%%%%%%%%%
In this paper, we are concerned with the \emph{chemical reaction network} (\emph{CRN}) model, which is frequently used in molecular programming~\cite{bPJ89,bFein19}.
CRNs are an abstraction of modern chemistry, Turing complete~\cite{cFLB+17}, and  deployable at the nanoscale with motifs such as DNA strand displacement~\cite{jSSW10,jCDS+13,cBSJ+17}.
Abstract molecule types in CRNs are called \emph{species} and are denoted with capital Roman characters such as \( A \), \( B \), and \( C \) and other decorations such as \( X_1 \), \( X_2 \), and \( Y^+ \), \( Y^- \).
Although there are many variations of the CRN model, here we use CRNs under deterministic mass action semantics since they are intrinsically \emph{analog}.
These deterministic CRNs are similar to other analog devices such as electrical circuits and Shannon's general purpose analog computer (GPAC)~\cite{jShan41,jGCB08c} and consist of systems of polynomial differential equations.

Formally, a \emph{chemical reaction network} (\emph{CRN}) \( N \) is a finite collection of \emph{reactions} of the form
\begin{equation}
	\rho = X_1 + X_2 + \cdots + X_n \goesto{k} Y_1 + Y_2 + \cdots + Y_m.
\end{equation}
Here the species \( X_1, \ldots, X_n \) are the \emph{reactants}, the species \( Y_1,\ldots,Y_m \) are the \emph{products}, and \( k \) is the \emph{rate constant} of the reaction \( \rho \).
Intuitively, a reaction specifies a relationship between molecular species, and in particular, how reactants combine to form products.
It is important to note that reactants and products may not be unique; for example, \( X + X + Y \goesto{1} X + X + X \) is a valid reaction and is commonly written \( 2X + Y \goesto{1} 3X \) for convenience.
The \emph{net effect} of a reaction \( \rho \) on a species \( X \), written \( \Delta\rho(X) \), is the difference of the multiplicities of \( X \) in \( \rho \)'s products and reactants.
For example, the net effect of \( 2 X + Y \goesto{1} Z \) on \( X \), \( Y \), and \( Z \) is -2, -1, and 1, respectively.
If a reaction has a net effect of zero on a reactant \( X \), then \( X \) is called a \emph{catalyst} of \( \rho \).

We now describe the semantics of deterministic chemical reaction networks under mass action kinetics. 
Let \( N \) be a CRN consisting of a finite set of reactions \( R \) over the species \( X_1, \ldots, X_n \).
Then \( N \) induces a polynomial initial value problem (PIVP) \( \mathbf{x} = (x_1, x_2, \ldots, x_n) \) where each variable \( x_i \) represents the real-valued concentration of the species \( X_i \).
Each variable \( x_i \) in the PIVP obeys the polynomial ordinary differential equation
\begin{equation}\label{eq:mass_action_odes}
	\D{x_i} = \sum_{\rho\in R}\Delta\rho(X_i)\cdot\text{rate}_\rho(t)
\end{equation}
where \( \text{rate}_\rho(t) \) is the rate of reaction \( \rho \) at time \( t \), defined to be the product of its rate constant along with each of its reactants.
Providing initial concentrations \( \mathbf{x}(0) = \mathbf{x}_0 \), the PIVP yields a unique solution \( \mathbf{x}(t) \).

As an example, consider the CRN defined by the reactions
\begin{align*}
	X + Y &\goesto{k_1} 2 Z\\
	X + Z &\goesto{k_2} 2 X\\
	Y + Z &\goesto{k_3} 2 Y.
\end{align*}
According to equation~\eqref{eq:mass_action_odes}, these reactions induce the ODEs
\begin{align*}
	\D{x} &= -k_1xy + k_2xz\\
	\D{y} &= -k_1xy + k_3yz\\
	\D{z} &= 2k_1xy - k_2xz - k_3yz.
\end{align*}

%%%%%%%%%%%%%%%%%%%%%%%%%%%%%%%%%%%%%%%%%%%%
% \subsection{Signal filters
% \label{sec:filters}}
%%%%%%%%%%%%%%%%%%%%%%%%%%%%%%%%%%%%%%%%%%%%
Electronic filters are used in many electronic devices, including radios, power lines, headphones, radar terminals, and many others.
Filters take an input signal and produce an output signal, and are often characterized by this input-output relationship.
This relationship is called a \emph{transfer function} and is simply the output divided by the input.
In linear systems, the output is related to the input through a linear differential equation and can be realized in electronic circuits with resistors, capacitors, and inductors.
One method of characterizing these filters utilizes the transfer function and Laplace transform to give a Bode plot, the response of the filter in terms of frequency.

\begin{figure}
	\centering
	\includegraphics[width=4.0in]{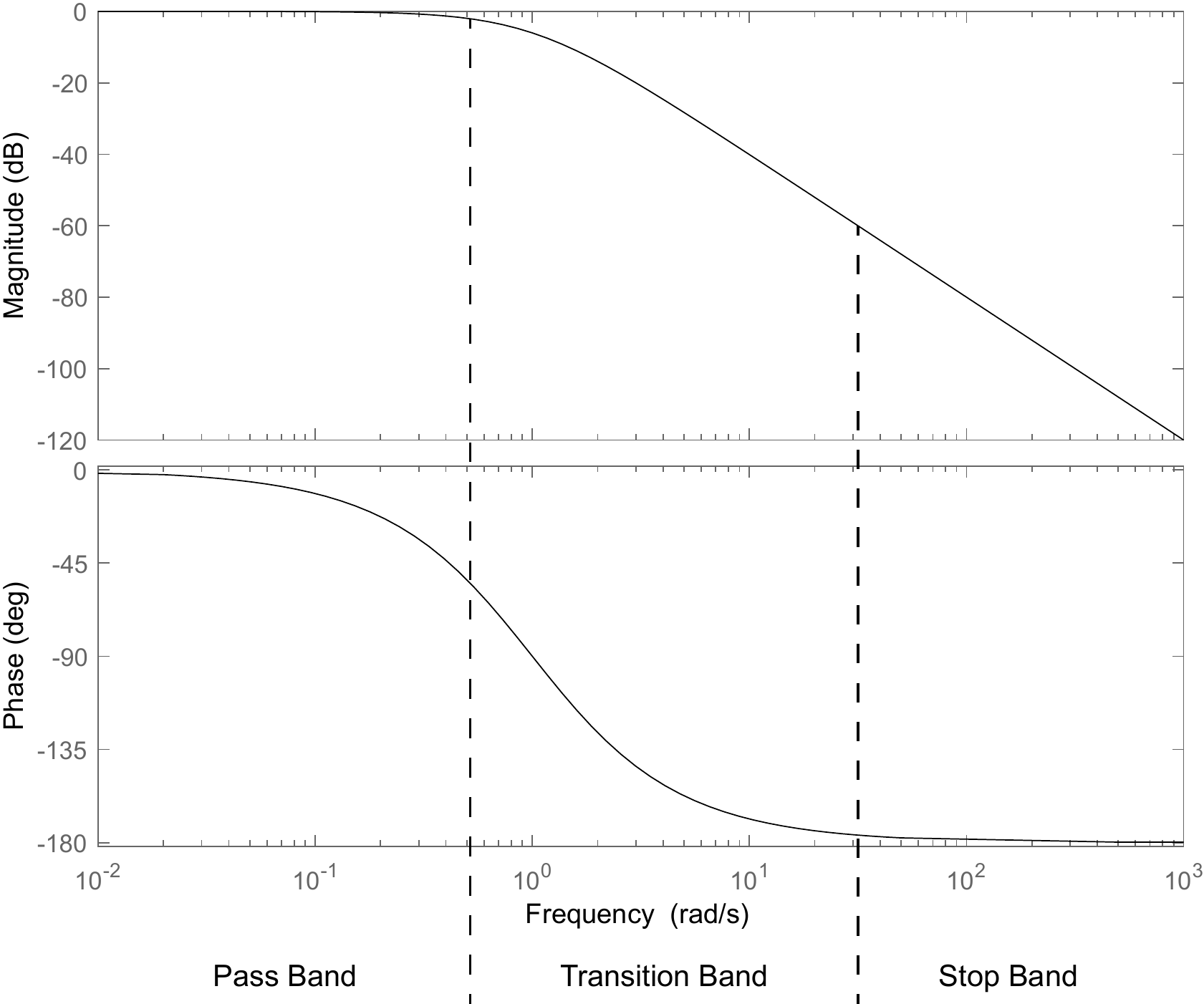}
	\caption{\label{fig:lowpassbodespec}
		Example Bode plot of a first-order low-pass filter, showing passband, transition band, and stopband at -60~dB.
	}
\end{figure}

Four common categories for filters include low-pass, high-pass, band-pass, and notch, characterized by how much input signal is transmitted at different frequencies.
For example, a low-pass filter transmits the input signal to the output at low frequencies but attenuates the input signal at higher frequencies.
This is depicted by a graph that shows the ratio of the output voltage to the input voltage with respect to frequency.
This ratio is measured using dB, \( 20 \log \frac{\text{output}}{\text{input}} \).
For example, if the output signal is half that of the input at a particular frequency, then it is approximately -6~dB lower.
If the output signal is double the input at a frequency, then the filter has a \emph{gain} (at that frequency), and in this case, it is a gain of 6~dB.
In addition to the magnitude of the output to the input, it is also necessary to define the phase change from the output to the input at each frequency.
These two graphs taken together are commonly referred to as a Bode plot of the response of the filter.
Figure~\ref{fig:lowpassbodespec} shows a low-pass filter with transfer function given as
\[
	\frac{Y(s)}{X(s)} = \frac{10}{10s + 10},
\]
where \( Y(s) \) is the input and \( X(s) \) is the output in the frequency domain.

\begin{figure}
	\centering
	\includegraphics[width=4.0in]{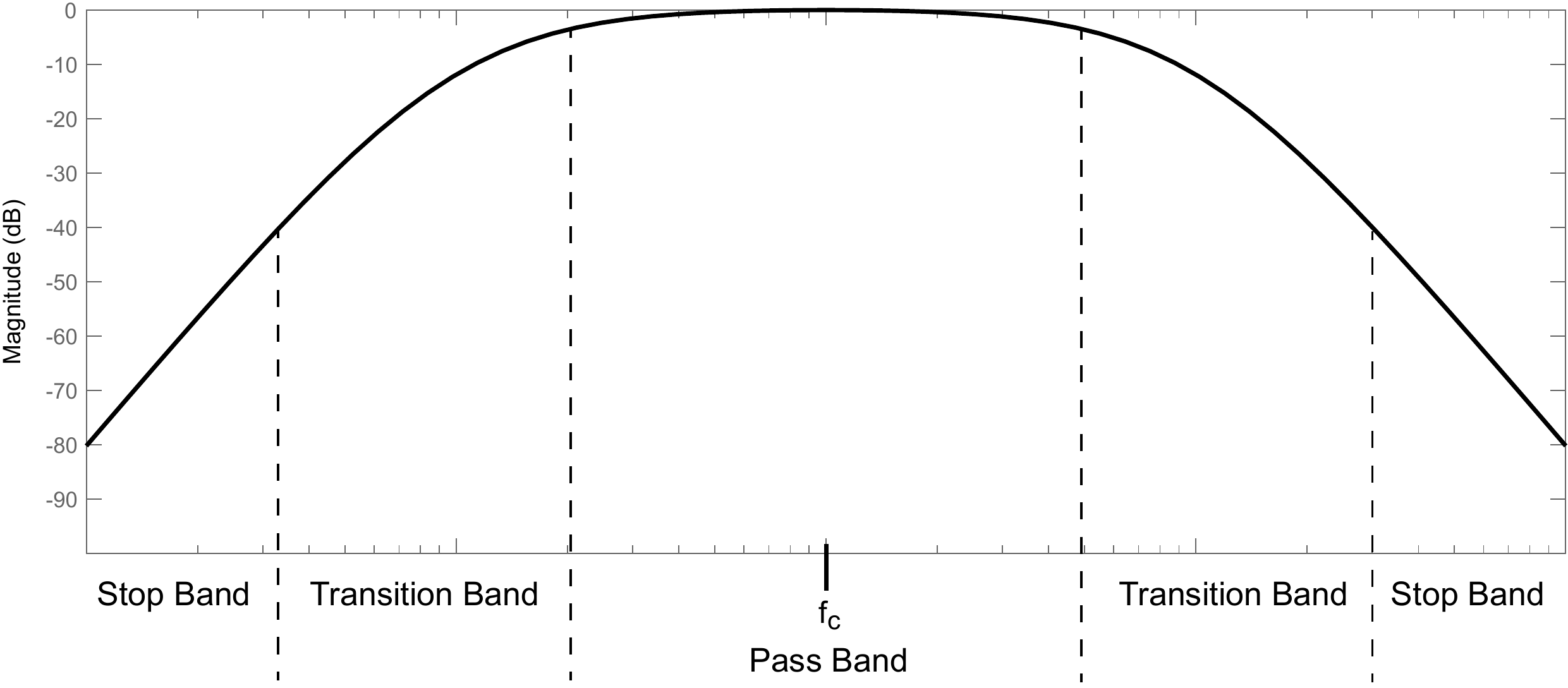}
	\caption{\label{fig:bandpassbodespec}
		Bode plot of a band-pass filter showing the passband, transition band, and stopband regions.
		\( f_c \) labels the center frequency of the passband region.
		For this filter, the stopband is 40~dB below the input signal.
		Note that the phase plot is not shown is this figure.}
\end{figure}

Besides characterizing a specific filter, Bode plots are also used to specify the requirements for filters.
For a low-pass filter, there are several critical parameters.
The parameters pertinent to our discussion here are described below.
\begin{enumerate}
	\item \textbf{Passband.}
	The passband specifies the region of frequencies that transmit the input signal to the output signal, and is typically specified by a single frequency \( f_c \) that defines the highest frequency where the output is above -3 dB of the input.

	\item \textbf{Stopband.}
	The stopband specifies the frequency range below an acceptable level of input leakage to the output.

	\item \textbf{Transition band.}
	The transition band is the range of frequencies between the passband and the stopband.
\end{enumerate}
Figure~\ref{fig:lowpassbodespec} depicts a low-pass filter specification where the stopband is defined as 60~dB below the input signal level.
A specification for a band-pass filter is shown in Figure~\ref{fig:bandpassbodespec}.
Here the center frequency of the passband is denoted \( f_c \), and the \emph{band width} of the filter is the range of frequencies that give an output signal that is above -3~dB of the input signal.
The high-pass filter is analogous to the low-pass filter except that the stopband is at low frequencies, and the passband is at high frequencies.
A notch filter is similarly an inverted band-pass filter that rejects (stop signals) in a specified range of frequencies.
Thus, these filters can also easily be specified in terms of Bode plots.

    %%%%%%%%%%%%%%%%%%%%%%%%%%%%%%%%%%%%%%%%%
\section{A programmable band-pass filter}
\label{sec:band_pass}
%%%%%%%%%%%%%%%%%%%%%%%%%%%%%%%%%%%%%%%%%
In this section, we describe a programmable band-pass filter and a natural implementation using chemical reaction networks.
By ``natural,'' we mean that the reaction network is small, straightforward, and does not approximate inputs, components, or functions of inputs.
In effect, the transfer function for the band-pass filter directly follows from the differential equations derived from a simple CRN.
Our construction will proceed by first constructing a natural low-pass filter.
This low-pass filter implementation is integral to the band-pass filter construction and demonstrates why high-pass filters cannot be implemented without approximation.

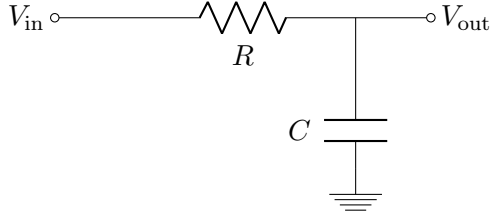
\begin{figure}
	\centering
	\begin{circuitikz}[american voltages]
		\draw
			(0,3) node[anchor=east] {$V_\text{in}$}
			(0,3) to [short, o-] (1,3)
			to [R, l_=$R$] (4,3)
			to [short, -o] (5,3)
			(5,3) node[anchor=west] {$V_\text{out}$};
		\draw
			(4,3) [short, -] (4,3)
			to [short, -] (4,2)
			to [C, l_=$C$] (4,1)
			to [short, -] (4,1)
			to (4,1) node[ground]{};
%			to [open, v^>=$V_IN$] (0, 6)
	\end{circuitikz}
	\caption{\label{fig:eelowpass}Circuit diagram of a low-pass filter}
\end{figure}

It is well known that a simple electrical low-pass filter can be constructed using a resistor and capacitor, as depicted in Figure~\ref{fig:eelowpass}.
We construct the chemical reaction equivalent of this electrical circuit by reverse-engineering the CRN from the ordinary differential equations (ODEs) given by Kirchhoff circuit laws and the low-pass circuit, which are shown below.
\begin{align}
    iR &= \vin - \vout  \label{eq:kirchhoff01}\\
    i  &= C \D{\vout}   \label{eq:kirchhoff02}
\end{align}
Substituting equation~\eqref{eq:kirchhoff02} into equation~\eqref{eq:kirchhoff01} yields the first-order ordinary differential equation:
\begin{equation}
    \D{\vout} = \frac{\vin - \vout}{RC}.
\end{equation}
We can convert this ODE directly into CRN~\ref{crn:simplelowpass} with two species \( \Vin \) and \( \Vout \), representing the two voltages.
This construction is commonly called \emph{pure pursuit} because one signal is ``pursuing'' the other.
In this case, the output species \( \Vout \) is ``chasing'' the input species \( \Vin \), and the rate constant of \( \frac{1}{RC} \) determines how fast \( \Vout \) chases \( \Vin\).
It is easy to verify that this CRN implements a low-pass filter described in Section~\ref{sec:prelim} with cutoff frequency \( f_c = 1/RC \) and gain factor \( k = 1/RC \).
We decouple the gain factor from the cutoff frequency by decoupling the rate constants in the CRN.
This is achieved by CRN~\ref{crn:lowpasswithgain} where the output concentration chases the input concentration times the gain factor \( k \).
This CRN's input/output behavior is exactly the standard transfer function for a first-order low-pass filter
\begin{equation}\label{eq:firstorderlowpassxfer}
	H(s) = \frac{k}{s + c},
\end{equation}
where $c$ is the cutoff frequency and $k$ is the gain.

\begin{figcrn}{Simple low-pass filter as pure pursuit}{crn:simplelowpass}
    \Vout &\goesto{\frac{1}{RC}} \emptyset   \\ 
    \Vin &\goesto{\frac{1}{RC}} \Vin + \Vout  
\end{figcrn}

\begin{figcrn}{Simple low-pass filter with gain}{crn:lowpasswithgain}
    \Vout &\goesto{c} \emptyset\\
    \Vin &\goesto{k} \Vin + \Vout
\end{figcrn}

The standard implementation of a band-pass filter is to compose a low-pass filter with a high-pass filter.
Unfortunately, the standard implementation of a high-pass filter using circuits yields a differential equation in which the derivative of the output voltage is dependent on the derivative of the input voltage.
For example, 
\begin{equation}
    \D{\vout} = \D{\vin} - \frac{\vout}{RC},
\end{equation}
is the differential equation derived from the standard first-order high-pass filter.
Although this is easily implemented as an electrical circuit using a capacitor, a CRN cannot compute the derivative of an arbitrary input signal exactly, unless the input signal can be anticipated and hard-coded into the CRN construction.

Approximating circuit behavior with CRNs has been studied in~\cite{jCTT20}.
In this paper we are interested in producing a band-pass filter with an arbitrary input without approximation, over all time, in a relatively small CRN.
Somewhat serendipitous, a simple approximation of the derivative leads directly to the band-pass filter that we desire.
A simple approximation of a derivative involves computing the difference between the input signal at two different times; unfortunately, creating a perfect time-delayed signal is not possible with a CRN.
Nevertheless, we can crudely approximate a time-delayed signal by using pure pursuit in the same way as described in the low-pass filter above.
It is then possible to approximate the derivative by subtracting the pursuing signal from the original, which gives a crude estimate of the rate of change over a short time interval.
Using this approximation, the resulting CRN induces a differential equation and transfer function that \emph{exactly} correspond to a second-order band-pass filter \emph{without approximation}.
Motivated by this, we consider the parameterized transfer function for a second-order filter shown below.
\begin{equation}\label{eq:secondorderxfer}
    H(s) = \frac{as}{s^2 + bs + c}
\end{equation}
Letting \( x \) represent the input signal and \( y \) represent the output signal, we have the corresponding differential equation
\[
	\frac{dy}{dt} = ax - by -cz,
\]
where \( z(t) \) is the function \( z = \int y(t) dt \).
We can now realize a CRN for this differential equation.
However, the \( -cz \) term in this ODE cannot be realized directly by a CRN, so we employ a construction introduced by Fages~\etal that implements such terms using the difference of two species~\cite{cFLB+17}.
This is commonly called the \emph{difference construction} and requires splitting each variable into two parts: \( x^+ \), \( x^- \), \( y^+ \), \( y^- \), \( z^+ \), and \( z^- \).
For this technique to work, it is critical to maintain the invariants:
\begin{align*}
    x(t) &= x^+(t) - x^-(t),\\
    y(t) &= y^+(t) - y^-(t),\\
    z(t) &= z^+(t) - z^-(t).
\end{align*}
Using this technique, we transform these ODEs into the following equivalent system of ODEs
\begin{align*}
    \frac{dy^+}{dt} - \frac{dy^-}{dt} &= a(x^+ - x^-) -b(y^+ - y^-) - c(z^+ - z^- ) \\
    \frac{dz^+}{dt} - \frac{dz^-}{dt} &= z^+ - z^-.
\end{align*}
We can now generate the corresponding CRN given below.
\begin{crn}
    X^+ &\goesto{a} X^+ + Y^+ \\
    X^- &\goesto{a} X^- + Y^- \\
    Y^+ &\goesto{b} \emptyset \\
    Y^- &\goesto{b} \emptyset \\
    Z^+ &\goesto{c} Z^+ + Y^- \\
    Z^- &\goesto{c} Z^- + Y^+ \\
    Y^+ &\goesto{1} Y^+ + Z^+ \\
    Y^- &\goesto{1} Y^- + Z^-
\end{crn}
Note that we also add the reactions
\begin{crn}
    Y^+ + Y^- \goesto{1} \emptyset\\
    Z^+ + Z^- \goesto{1} \emptyset
\end{crn}
to bound the concentrations of \( Y^+, Y^-, Z^+ \), and \( Z^- \), but this does not affect the solution to the differential equations.
The rate constants \( a \), \( b \), and \( c \) can be replaced by catalyzed biomolecular reactions with unity rate constants.
This gives us the ability to tune the filter using concentrations of species rather than using rate constants in the reactions.

\begin{figure}
    \begin{minipage}{2.5in}
    \begin{crn}
        A + X^+ &\goesto{1} A + X^+ + Y^+ \\
        A + X^- &\goesto{1} A + X^- + Y^- \\
        B + Y^+ &\goesto{1} B \\
        B + Y^- &\goesto{1} B \\
        C + Z^+ &\goesto{1} C + Z^+ + Y^- \\
        C + Z^- &\goesto{1} C + Z^- + Y^+ \\
        Y^+ &\goesto{1} Y^+ + Z^+ \\
        Y^- &\goesto{1} Y^- + Z^- \\
        Y^+ + Y^- &\goesto{1} \emptyset\\
        Z^+ + Z^- &\goesto{1} \emptyset
    \end{crn}
    \end{minipage}
    \begin{minipage}{2.5in}
        \includegraphics[width=2.5in]{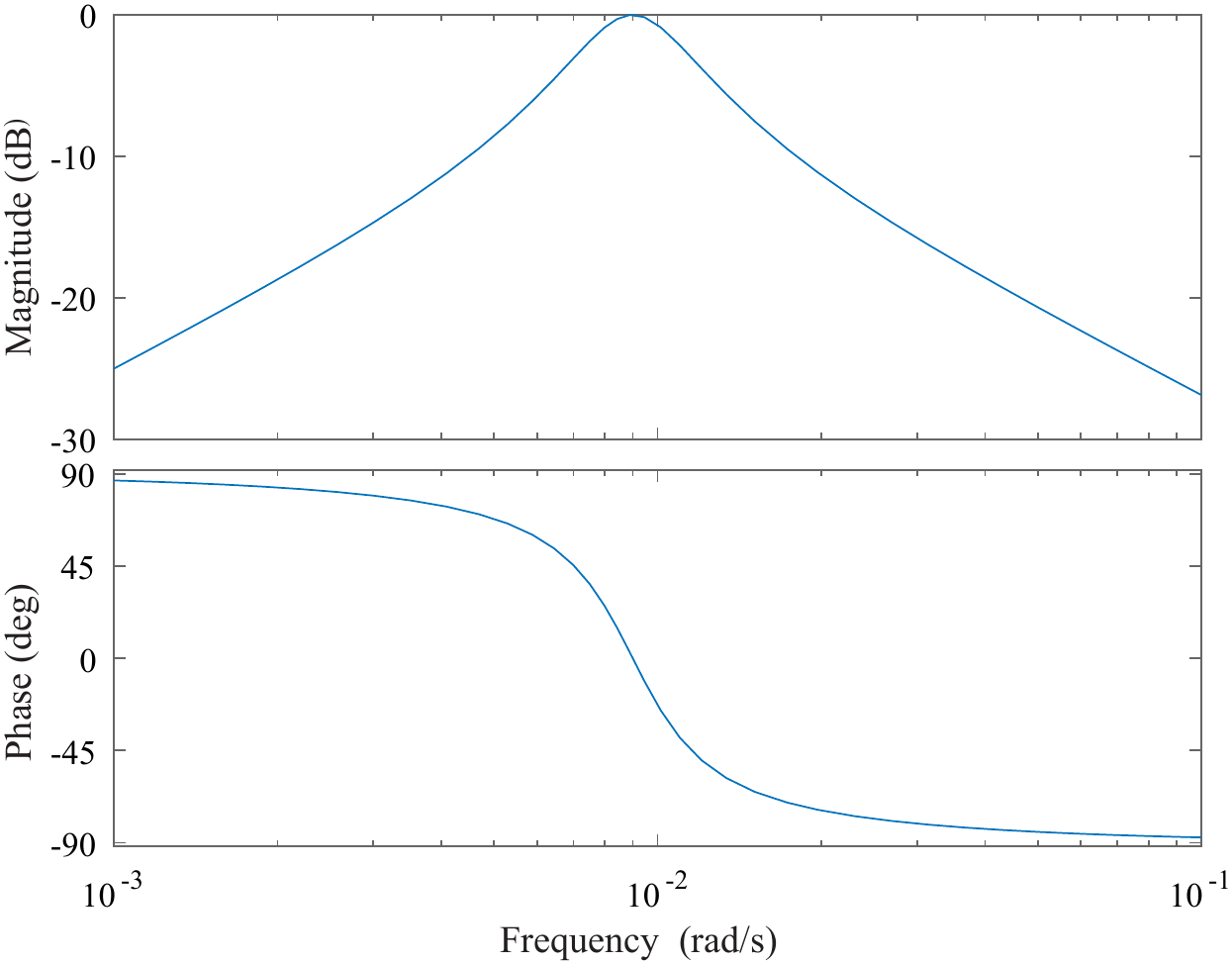}
    \end{minipage}
    \caption{\label{fig:exampleband-pass}
        Band-pass filter with center frequency \( 0.009 \) rads/sec, bandwidth \( 0.0045 \) rads/sec, and gain of \( 1 \), (\( a = 0.0045 \), \( b= 0.0045 \), \( c = 0.000081 \))
    }
\end{figure}

\begin{figure}
    \centering
    \includegraphics[width=2in]{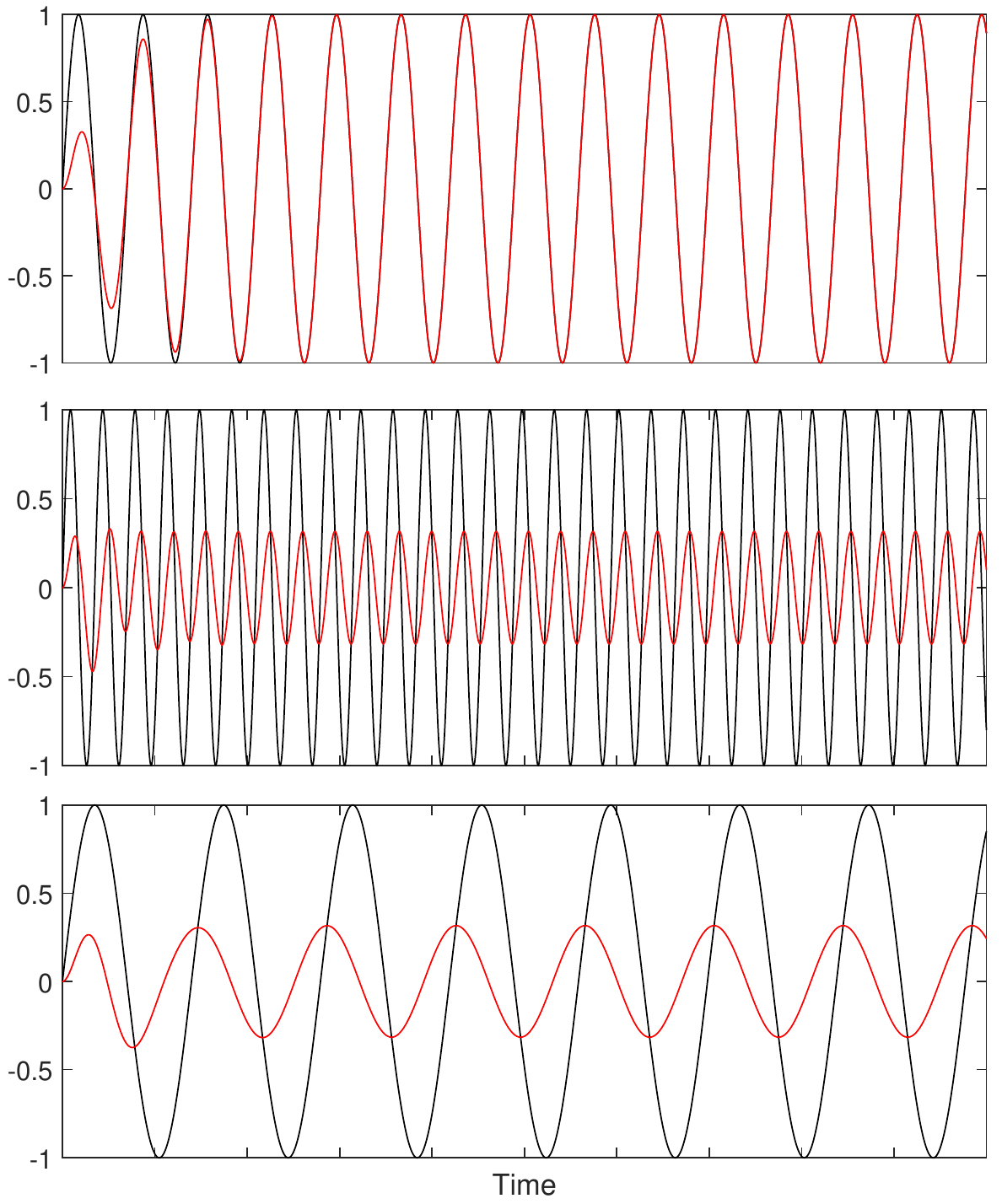}
    \caption{\label{fig:band-passmatlabsim}
        Matlab simulation of band-pass filter.
        Input signal is depicted in black, output signal depicted in red.
        Top graph input is sin wave at band-pass center frequency of 0.009 rads/sec.
        Middle graph is twice the venter frequency, and bottom is half the center frequency.
    }
\end{figure}

The Laplace transform of a second-order band-pass filter transfer function in terms of gain \( k \), quality \( Q \), and center frequency \( \omega_0 \) is given by
\[
    H(s) = \frac{k(\omega_0/Q)s}{s^2+(\omega_0/Q)s + \omega_0^2}.
\]
By assigning \( a \), \( b \), and \( c \) appropriately, we can realize any second-order band-pass filter of this form in a CRN with no approximation over all time.
Thus we have that 
\begin{align*}
	a = &k(\omega_0/Q) \\
	b = &\omega_0 / Q \\
	c = &\omega_0^2.
\end{align*}
For example, Figure~\ref{fig:exampleband-pass} shows the CRN implementation and Bode plots of a band-pass filter with gain \( 1 \), center frequency \( 0.009 \) radians per second, and bandwidth \( 0.0045 \) radians per second.
Figure~\ref{fig:band-passmatlabsim} shows a Matlab simulation of this band-pass filter with three different input signal frequencies:
(1)~exactly the center frequency (0.009 rads/sec),
(2)~twice the center frequency, and
(3)~half the center frequency.
As expected, the output signal of the filter with input exactly the center frequency outputs the signal at unity gain.
The output of the other two frequencies are appropriately reduced.

    %%%%%%%%%%%%%%%%%%%%%%%%%%%%%%%%%%%%%%%%%%%%%%%%%%%%%
\section{Modulation/demodulation of chemical signals}
\label{sec:modulation}
%%%%%%%%%%%%%%%%%%%%%%%%%%%%%%%%%%%%%%%%%%%%%%%%%%%%%
Encoding and transmitting many signals through a single medium is utilized in a variety of applications, including AM and FM radio, cable TV,  ADSL, and others.
In this section, we describe how a chemical reaction network may be designed to encode and decode amplitude modulated (AM) chemical concentrations signals.
We further describe a method where these concentration signals can also be encoded using frequency modulation (FM).

%%%%%%%%%%%%%%%%%%%%%%%%%%%%%%%%%
\subsection{Amplitude modulation}
%%%%%%%%%%%%%%%%%%%%%%%%%%%%%%%%%
Amplitude modulation (AM) encodes a signal on a \emph{carrier} frequency by modulating the amplitude of the wave proportional to the signal.
Due to the superposition principal, multiple signals can be encoded using different carrier frequencies and then transmitted over a single medium.
In the case of chemical concentrations, our goal is to transmit and receive multiple signal concentrations sent through a single species that encodes all the transmitted signals.

It is relatively straight-forward to encode an input signal with amplitude modulation.
The input signal \( u(t) \) is multiplied by a carrier, \( \sin(\theta t) \) where \( \theta \) is the \emph{carrier frequency}.
Thus, implementing an AM modulator in a chemical reaction network requires (1)~generating the carrier \( \sin(\theta t) \), and (2)~multiplying the input signal \( u(t) \) by the carrier signal.

We begin with the CRN construction for generating the carrier signal which consists of the reactions:
\begin{crn}
	F + \Cp   &\goesto{} F + \Cp + \Sp\\
	F + \Cm   &\goesto{} F + \Cm + \Sm\\
	F + \Sm   &\goesto{} F + \Sm + \Cp\\
	F + \Sp   &\goesto{} F + \Sp + \Cm\\
	\Sp + \Sm &\goesto{} \emptyset\\
	\Cp + \Cm &\goesto{} \emptyset
\end{crn}
The above CRN is designed in a \emph{dual rail} scheme such that
\begin{align*}
	\sp(t) - \sm(t) &= \sin(ft)\\
	\cp(t) - \cm(t) &= \cos(ft)
\end{align*}
is satisfied for all \( t \ge 0 \) as long as \( \sp(0) - \sm(0) = 1 \) and \( \cp(0) - \cm(0) = 0 \) is satisfied.
For convenience, we define the function \( s \) by \( s(t) = \sp(t) - \sm(t) \) for all \( t \ge 0 \).
It is important to observe that the net effect on the species \( F \) is zero.
By design, \( F \) is constant and serves as a means of tuning the carrier frequency of the sine wave, akin to tuning an AM radio.
\( F \) can also be changed dynamically while the CRN is active.

\begin{figure}
	\begin{center}
		\includegraphics[width=2.5in]{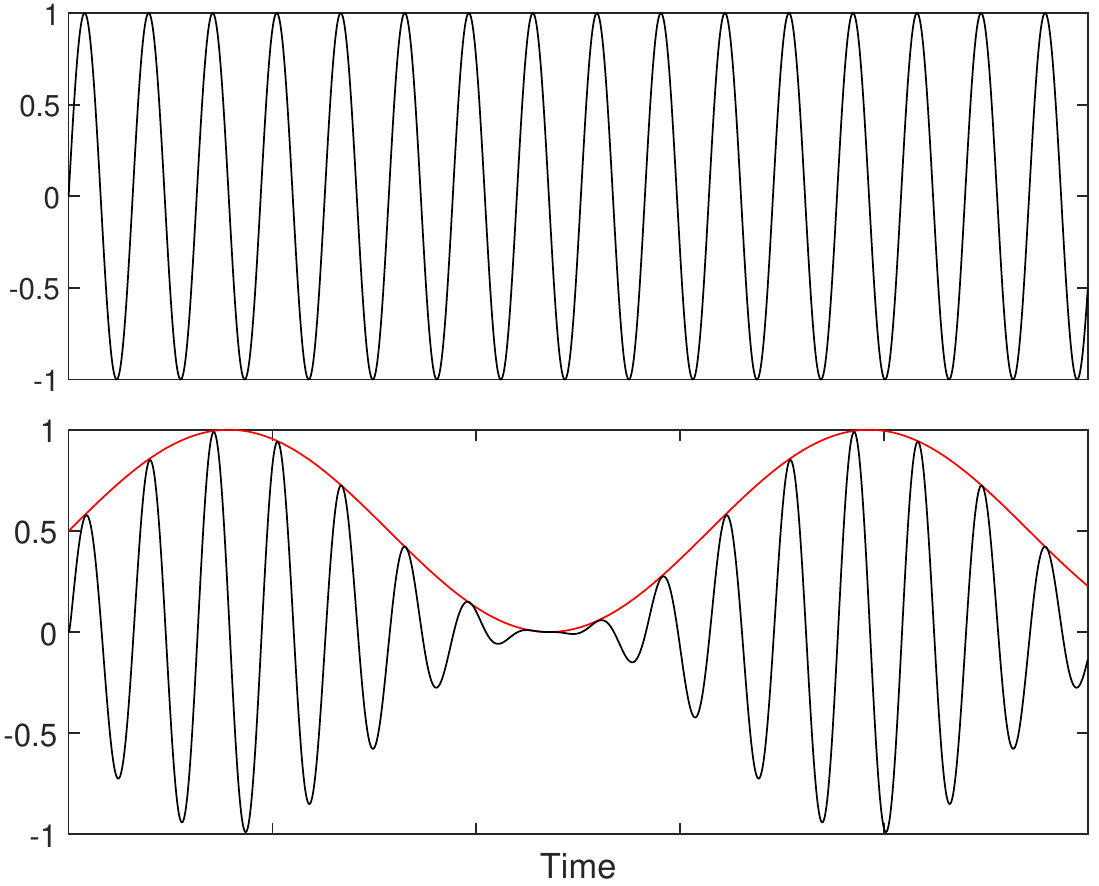}
		\caption{\label{fig:singlemodulationsim}
			Single signal modulation with carrier frequency of \( 0.1 \) rads/sec and a simple sine wave signal of \( 0.01 \) rads/sec.
			The top graph is the carrier signal and the bottom graph is the signal superimposed over the modulated signal.
		}
	\end{center}
\end{figure}

With the carrier signal generated via \( s = \sp - \sm \), it remains to compute the modulated signal from the input signal by a simple multiplication.
Our approach uses the pure pursuit technique described in Section~\ref{sec:band_pass} to approximate an exact multiplication of the two signals.
The ODE for this approximation is
\[
	\D{m} = us - m
\]
where \( m = \mp - \mm \) is also a dual-rail signal.
In this case, the value of \( m(t) \) is ``pursuing'' the value of the product \( u(t)\cdot s(t) \).
In effect, this is a low-pass filter with the input signal modulating the gain.
The approximation can be improved by uniformly increasing the rate constants of the reactions below.
\begin{crn}
	\Mp       &\goesto{} \emptyset\\
	\Mm       &\goesto{} \emptyset\\
	U + \Sp &\goesto{} U + \Sp + \Mp\\
	U + \Sm &\goesto{} U + \Sm + \Mm
\end{crn}
It is easy to verify that the reactions above induce the ODEs
\begin{align*}
	\D{\mp} - \D{\mm}
		&= \left(-\mp + u\sp\right) - \left(-\mm + u\sm\right) \\
		&= u\left(\sp - \sm\right) - \left(\mp - \mm\right)\\
		&= us - m.
\end{align*}
Figure~\ref{fig:modulation} defines the complete CRN that modulates a signal on a specified carrier frequency and Figure~\ref{fig:singlemodulationsim} is a Matlab simulation of this CRN.
In this simulation, the signal to modulate is a simple sine wave that is one tenth the frequency of the carrier wave.

\begin{figcrn}{Modulation CRN}{fig:modulation}
		F + \Cp   &\goesto{} F + \Cp + \Sp\\
	F + \Cm   &\goesto{} F + \Cm + \Sm\\
	F + \Sm   &\goesto{} F + \Sm + \Cp\\
	F + \Sp   &\goesto{} F + \Sp + \Cm\\
	\Sp + \Sm &\goesto{} \emptyset\\
	\Cp + \Cm &\goesto{} \emptyset \\
		\Mp       &\goesto{} \emptyset\\
	\Mm       &\goesto{} \emptyset\\
	U + \Sp &\goesto{} U + \Sp + \Mp\\
	U + \Sm &\goesto{} U + \Sm + \Mm
\end{figcrn}

It is easy to combine the modulation scheme described above for \( n > 1 \) signals \( u_1(t), \ldots, u_n(t) \) with corresponding carriers \( s_1(t), \ldots, s_n(t) \) and transfer all of them simultaneously through the single dual-railed signal \( m(t) \).
This is accomplished by summing all of the modulated signals \( m(t) \) via a CRN that approximate the sum \( \sum_{i=1}^{n}u_i(t)\cdot s_i(t) \).
The CRN for this approximation is shown in CRN~\ref{crn:superposition}.
It is easy to verify that the ODE for \( m(t) \) is
\[
	\D{m} = \sum_{i=1}^{n}u_i\cdot s_i - m,
\]
which can again be made arbitrarily precise by increasing the rate constants.

\begin{figcrn}{Superposition of all modulated signals}{crn:superposition}
	\Mp       &\goesto{} \emptyset\\
	\Mm       &\goesto{} \emptyset\\
	U_1 + \Sp_1 &\goesto{} U_1 + \Sp_1 + \Mp\\
	U_1 + \Sm_1 &\goesto{} U_1 + \Sm_1 + \Mm\\
	&\quad\;\vdots\\
	U_n + \Sp_n &\goesto{} U_n + \Sp_n + \Mp\\
	U_n + \Sm_n &\goesto{} U_n + \Sm_n + \Mm.
\end{figcrn}

\begin{figure}
	\begin{center}
		\includegraphics[width=2.5in]{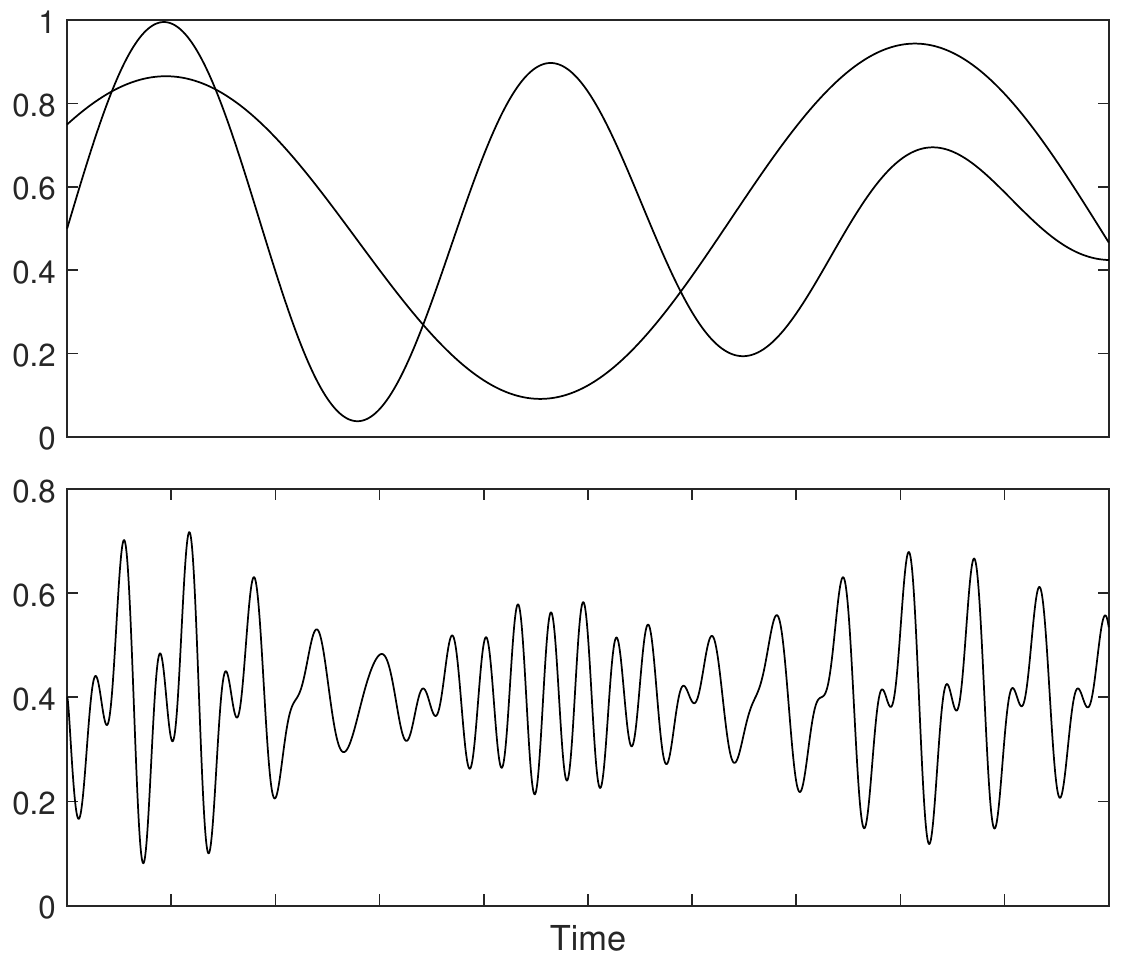}
		\caption{\label{fig:doublemodulation} Two signals are shown in the top graph.
		The combined modulated signal is shown in the bottom graph.}
	\end{center}
\end{figure}

Note that only \emph{one} copy of the reactions \( \Mp \rightarrow \emptyset \) and \( \Mm \rightarrow \emptyset \) are necessary even though there are \( n \) signals being passed through \( m(t) \).
Figure~\ref{fig:doublemodulation} shows two signals that are modulated at two different carrier frequencies, 0.1 rads/sec and 0.2 rads/sec.
The combined superimposed signal is also shown.

%%%%%%%%%%%%%%%%%%%%%%%%%%%%%%%%%%%
\subsection{Amplitude Demodulation}
\begin{figure}
	\begin{center}
	\begin{tikzpicture}
	\node[dspnodeopen,dsp/label=left] (c00){Input \( m \)};
	\node[dspfilter,right= of c00,
		 minimum width=2.5cm, 
		 minimum height=1.5cm,
		 text height=2em]       (c0) {Band-pass\\filter};
	\node[dspfilter, right= of c0,
		 minimum width=2.5cm, 
		 minimum height=1.5cm,
		 text height=2em]                    (c1) {Low-pass \\ filter};
	\node[dspnodeopen,right= of c1,dsp/label=right](c2){Output \( y \)};
	\draw[dspconn] (c00) -- (c0);
	\draw[dspconn] (c0) -- (c1);
	\draw[dspconn] (c1) -- (c2);
	\end{tikzpicture}
	\end{center}
	\caption{\label{fig:tunnerblock}
		Block diagram...
	}
\end{figure}
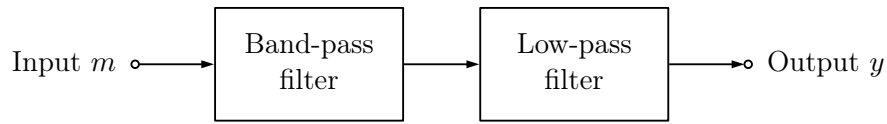

\begin{figure}
	\begin{center}
	\framebox{
	\includegraphics[height=7.0in]{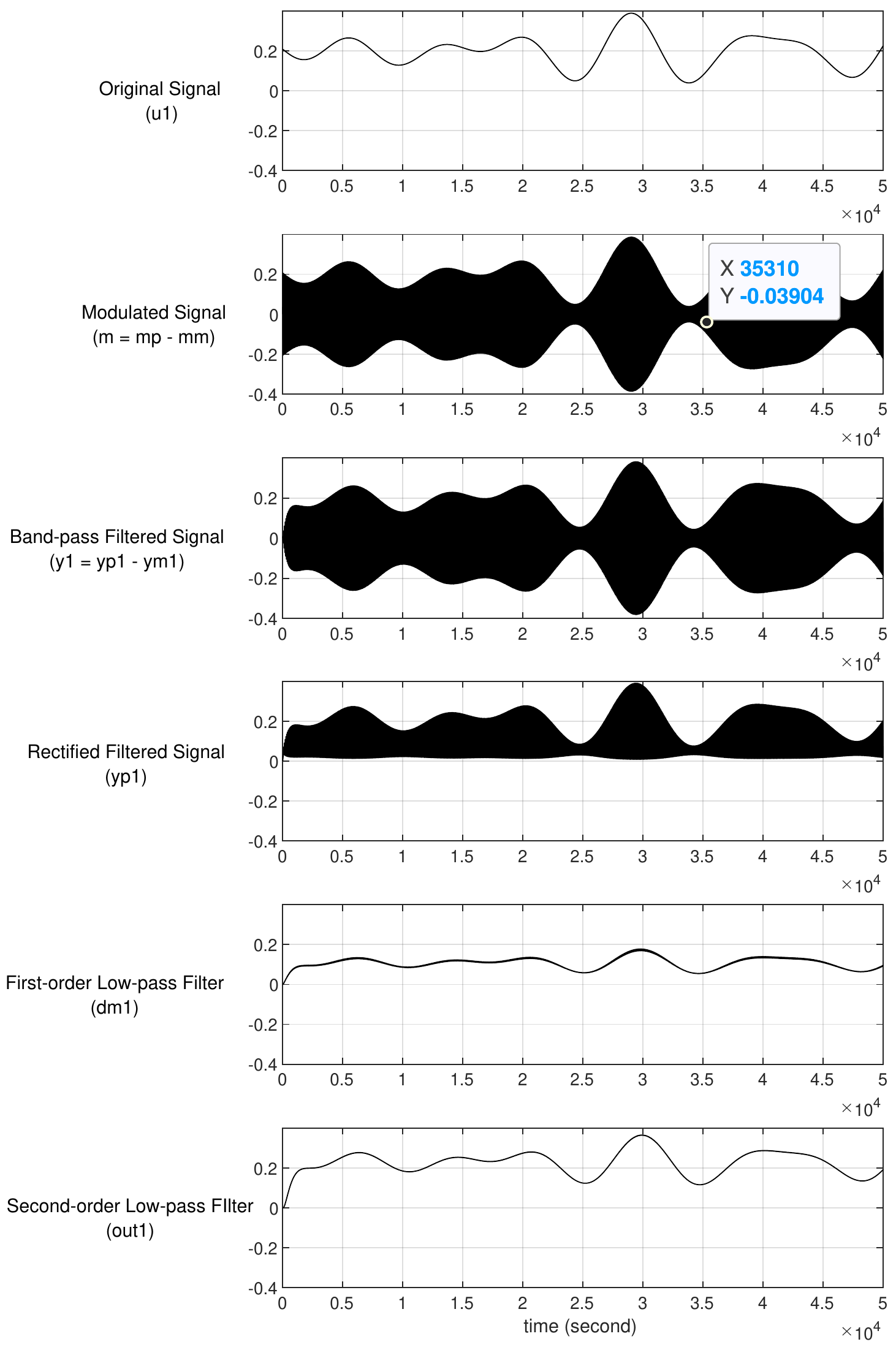}
}
	\caption{\label{fig:completemoddemo} Complete Mod demod...}
	\end{center}
\end{figure}

\begin{figure}
	\begin{center}
		\framebox{
	\includegraphics[height=7.0in]{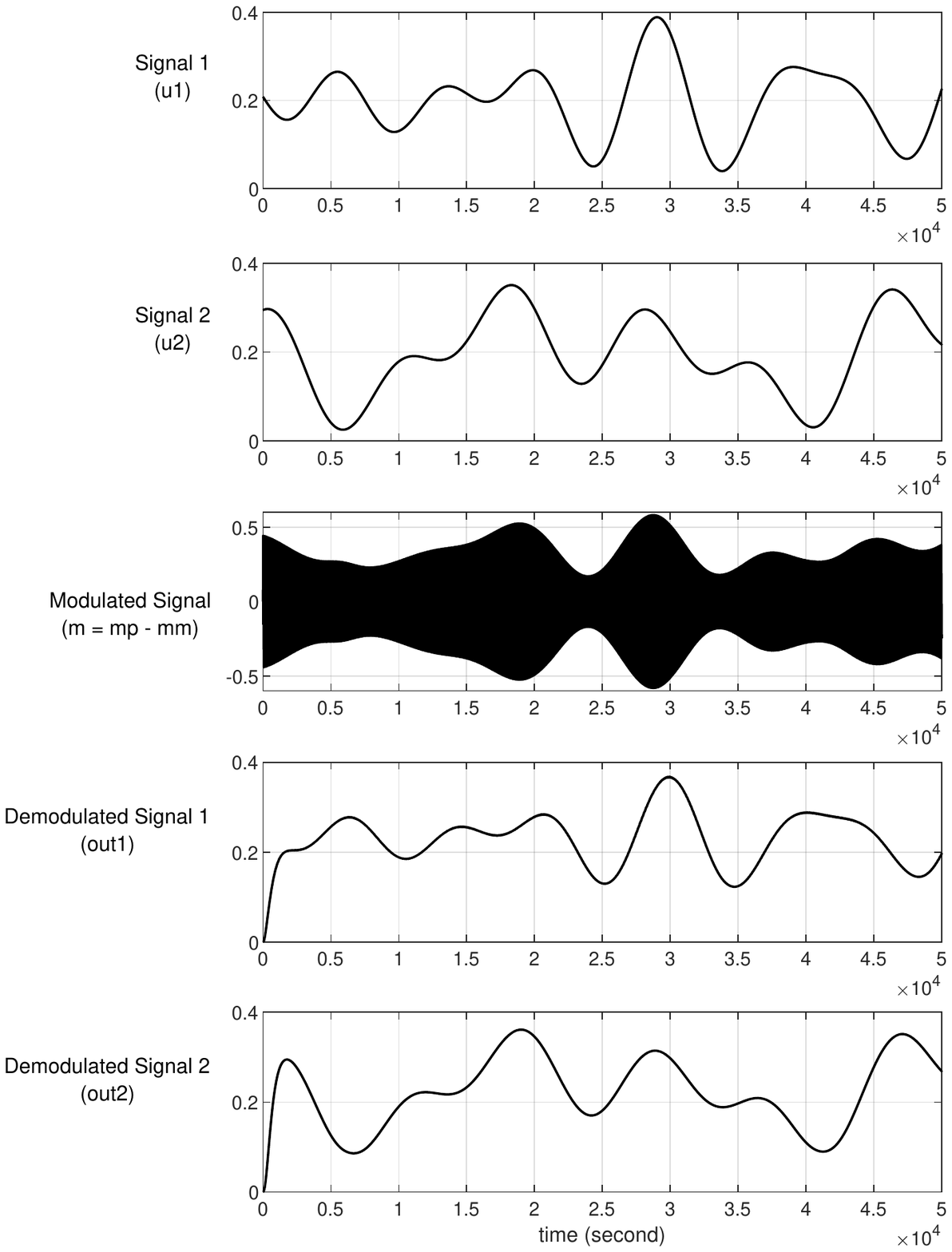}
}
	\caption{\label{fig:completemoddemo2} Multi Signals Example...}
	\end{center}
\end{figure}

%%%%%%%%%%%%%%%%%%%%%%%%%%%%%%%%%%%
We now describe a CRN that given a signal encoding many modulated signals encoded outputs an approximation of the original signal, similar to how an AM radio may be tuned to decode a signal for a specific radio station at a specific frequency.
A simple method for doing this utilizes a band-pass filter to select a specific carrier frequency to pass, followed by a low-pass filter on just the positive signal to remove the carrier frequency but leave the original signal.
The equivalent circuit is known as a diode detector, one of the simplest methods to demodulate amplitude modulated signals.
Figure~\ref{fig:tunnerblock} shows a high-level block diagram.
Both the band-pass and low-pass filters were discussed in Section~\ref{sec:band_pass}, thus it remains to compose them together and select the appropriate parameters.
In Figure~\ref{fig:completemoddemo}, we show an example of an input signal, modulated by a carrier frequency of 0.1 rads/sec, and then demodulated to give an output signal.
The Modulated Signal and the Band-pass Filtered Signal are very close to the same since there is only a single modulated signal.
It is very simple to rectify the Band-pass filtered signal so that only the positive portion remains by only utilizing the positive component \(yp1\) of \(y1\) shown in Figure~\ref{fig:completemoddemo}.
The bottom two graphs depict the output of the Rectified Filter signal after passing though a low-pass filter to recover the original signal.
Two different low-pass filters are demonstrated.
The first is a simple first-order filter, used when radio was first invented.
The second shows an improvement in the output when a second-order filters is used with some gain.
Finally in Figure~\ref{fig:completemoddemo2} we show two signals that are modulated on two different frequencies of 0.1 rads/sec and 0.2 rads/sec, their resulting modulated signal, and the demodulated output of the two signals.
It is worth noting again that this scheme is not limited two signals.

\subsection{Discussion}
This section shows that it is possible to amplitude modulate signals and then demodulate them using chemical reaction networks.
In fact, any number of signals can be encoded with carrier waves of different frequencies, summed together, and then transmitted through a single dual-railed species.
While the examples above utilize only first-order differential equations, higher performance filters are easily generated by composing first-order filters, or directly implementing a higher-order differential equation.
These lead to all of the standard low-pass and band-pass filters found in literature and textbooks, including Butterworth, elliptical, and Chebyshev filters.
However, any linear circuit may be implemented with chemical reaction networks, provided that the differential equation(s) do not require the derivative of an arbitrary input signal.
These filters operate identical to their circuit counterparts, and the performance does not degrade over time.
Note that it is possible to create a band-limited version of the high-pass filter using a band-pass filter, or a notch filter by composing two band-pass filters in series.

Frequency modulation (FM) is also used to transmit data over a carrier frequency, except that the frequency is modulated instead of the amplitude.
This modulation is easily accomplished with the CRNs above by using the sine wave generator to again create a carrier frequency.  However, instead of using a constant-value species $F$ to set the frequency, we instead add in the signal to transmit to the constant frequency, and then use this signal to dynamically set the frequency of the sin wave generator.
This gives the desired frequency modulated signal.
The demodulation of frequency modulated signals is accomplished with two band-pass filters and a low-pass filter.

    \bibliographystyle{plain}

\end{document}